\documentclass[aps,twocolumn]{revtex4}
\usepackage{graphics}

\begin{document}

\title{ Chiral soliton models, large $N_c$ consistency
and the $\Theta^+$ exotic baryon}

\author{Thomas D. Cohen}
\email{cohen@physics.umd.edu}

\affiliation{Department of Physics, University of Maryland, College
Park, MD 20742-4111}

\begin{abstract}
Predictions for a light collective $\Theta^+$ baryon state (with
strangeness +1) based on the collective quantization of chiral
soliton models are shown to be inconsistent with large $N_c$ QCD.
The lightest strangeness +1 state to emerge from the analysis has
an excitation energy which at large $N_c$ scales as $N_c^0$ while
collective quantization is legitimate only for excitations which
go to zero as $N_c \rightarrow \infty$.  This inconsistency
strongly suggests that predictions for $\Theta^+$ properties based
on collective quantization of chiral solitons are not valid.
\end{abstract}


\maketitle

 There has been considerable recent excitement in
hadronic physics.  Several experimental groups have announced the
identification of a narrow baryon resonance with a strangeness of
+1 ({\it i.e.} containing one excess strange antiquark)
\cite{exp}. Such a state is manifestly exotic in the sense of the
quark model---it cannot be a simple three-quark state. This
discovery has prompted considerable theoretical interest. Much of
the theory has been in the context of generalized quark models in
which the new baryon is identified as a
pentaquark\cite{StaRis,CapPagRob,Wyb,Hos,JafWil,Car,Che,Glo,Mat}.
Unfortunately, the nature of this analysis is highly model
dependent---there is no obvious way to see how phenomenological
quark models emerge from QCD---and thus probably should be
regarded presently as somewhat speculative. One theoretical
approach to the problem clearly stands out---the analysis based
on the SU(3) chiral soliton model treated with collective
quantization\cite{Pres,DiaPetPoly,WalKop,BorFabKob,Kim}. This
analysis has three obvious virtues: i) The calculation predates
the observation\cite{Pres,DiaPetPoly}; ii) it made a strikingly
accurate prediction of the mass\cite{Pres,DiaPetPoly} and has
predicted a narrow width\cite{DiaPetPoly} consistent with those
presently observed \cite{Nus}; and iii) although apparently based
on a particular model---the chiral soliton model---the analysis is
completely insensitive to the details of the model such as the
profile function which emerges from the detailed dynamics.

This third point is particularly important.  There has been
considerable experience over the years with relations in chiral
soliton models which are independent of the dynamical details
going back  nearly twenty years\cite{AdkNap}.  Typically such
relations are exactly satisfied in the large $N_c$ limit of QCD;
the relations are derivable directly from large $N_c$ consistency
relations\cite{GS,DM,Jenk,DJM}. This holds for relations of
typical static observables (such as magnetic moments or axial
couplings) considered in ref.~\cite{AdkNap} and also for more
esoteric quantities such as the nonanalytic quark mass dependence
of observables near the chiral limit\cite{CohBro,Coh} or
meson-baryon scattering observables \cite{CohLeb}. Thus, it seems
plausible that the analysis of
refs.~\cite{Pres,DiaPetPoly,WalKop,BorFabKob,Kim} is similarly
model-independent.

At first blush, this is quite satisfying: it appears that the
observed $\Theta^+$ state can be easily understood in terms of
large $N_c$ QCD and SU(3) flavor.  The issue addressed in this
paper is whether this is, in fact, true. Despite the remarkable
phenomenological success in predicting the mass and width of the
$\Theta^+$ seen in ref.~\cite{DiaPetPoly}, {\it a priori} there
is a compelling reason to doubt the validity of the analysis.
Surprisingly this reason is {\it not} that the predicted state is
a large $N_c$ artifact but is associated with a more basic issue
with soliton quantization.  Here it is found that the prediction
for the $\Theta^+$ arises due to an inconsistent implementation
of large $N_c$ scaling in the soliton model; the prediction is an
artifact of the treatment of collective variables in the model.
In particular, it is shown here that the prediction depends on
using collective quantization of the soliton outside the regime
of validity of this method: states with positive strangeness such
as the $\Theta^+$ necessarily have an excitation energy of order
$N_c^0$ while the semi-classical quantization method used to
predict the state is only valid for excitations of order
$N_c^{-1}$.  An alternative argument based on general features of
baryon states in large $N_c$ QCD also indicates that the
predicted $\Theta^+$ state is spurious.

Let us begin by briefly reviewing the essential aspects of the
analysis of refs.~\cite{Pres,DiaPetPoly,WalKop,BorFabKob,Kim}. The
starting point is a treatment of SU(3) chiral soliton models which
was developed in the mid-1980s\cite{SU3Quant}. In this approach
one finds a classical static ``hedgehog'' configuration in an
SU(2) subspace (the u-d subspace). The details of the profile are
model dependent but the general structure of the theory is not.
If one neglects SU(3) symmetry breaking effects then there are
eight collective (rotational) variables which are then quantized
semi-classically using an SU(3) generalization\cite{SU3Quant} of
the usual SU(2) collective quantization scheme\cite{ANW}.  The
collective Hamiltonian is given by
\begin{equation}
H^{\rm rot} = \frac{1}{2 I_1} \sum_{A=1}^3 {\hat{J}_A'}{}^2 \, +
\, \frac{1}{2 I_2} \sum_{A=4}^7 {\hat{J}_A'}{}^2   \; ,
\label{collective}
\end{equation} where $I_1$ ($I_2$) is the moment of inertia within
(out of) the SU(2) subspace and $\hat{J}_A'$ are generators of
SU(3) in a body-fixed (co-rotating) frame. Again, the numerical
values of the moments of inertia are model dependent but the
structure is not. There is an additional quantization  constraint
\begin{equation}
{J'_8}=-\frac{N_c B }{2\sqrt{3}} \;  ,\label{quantcond}
\end{equation}
where $B$ is the baryon number.

The explicit factor of $N_c$ in eq.~(\ref{quantcond}) plays a
central role in this paper and it is useful to understand its
origin. In Skyrme type models it follows directly from the
Witten-Wess-Zumino term (which topology fixes to be an integer
that can be identified with $N_c$). It can also be easily
understood at the quark level. In a body-fixed frame the baryon
number is associated with the SU(2) sub-manifold. There is also a
body-fixed hypercharge associated with this sub-manifold which is
related to the SU(3) generator in the usual manner: $Y'= -2
{J'}_8/ \sqrt{3}$.  There is a general relation relating the
baryon number, hypercharge and strangeness at large $N_c$ which
is valid at arbitrary $N_c$ :
\begin{equation}
Y = \frac{N_c B}{3} + S \label{hyper} \; \; ;
\end{equation}
this only coincides with the familiar relation $Y=B+S$ for
$N_c=3$. Equation (\ref{hyper}) follows from the fact that the
hypercharge of up, down and strange quarks as being 1/3, 1/3 and
-2/3, respectively. (These are the standard hypercharges of
quarks in an $N_c=3$ world.  These hypercharge assignments must
hold for general $N_c$ provided hypercharge is  isosinglet and
traceless in SU(3) and has the property that the hypercharge of
mesons is equal to the strangeness.) Given the fact that all three
flavors of quark all have baryon number of $1/N_c$ while the
strangeness is zero for u and d quarks and -1 for s quarks, one
sees that eq.~(\ref{hyper}) must hold. To complete the derivation
of eq.~(\ref{quantcond}), note that in a body-fixed frame, the
SU(2) sub-manifold has zero strangeness; accordingly
eq.~(\ref{hyper}) implies that $Y' = N_c B/3$ and the
quantization condition in eq.~(\ref{quantcond}) immediately
follows.

The masses which emerge from this depend on the quadratic Casimir
of the SU(3) multiplet, $C_2= \sum_{A=1}^8 \hat{J}_A^2$,  and the
angular momentum, $J$:
\begin{eqnarray}
M_{SU(3)} & = &  M_0 + \frac{C_2}{2 I_2} + \frac{(I_2 -
I_1) J (J+1) }{2 I_1 I_2} - \frac{N_c^2}{24 I_2} \; , \nonumber \label{mass} \\
{\rm with} \; \; &C_2 & =  \left( p^2 + q^2 + p q + 3(p +q)\right
)/3  \; ,
 \end{eqnarray}
 where $M_0$ is a common soliton mass.  $C_2$ is the quadratic Casimir
and is expressed  in terms of labels $p,q$ which denote the SU(3)
representation. The quantization condition in
eq.~(\ref{quantcond}) greatly restricts the possible SU(3)
representations: only SU(3) representations which contain
hypercharge equal to $N_c /3$ are allowed: if the hypercharge in
a body-fixed frame satisfies eq.~(\ref{quantcond}), the
representation will include a state with that hypercharge.
Moreover, since in the SU(2) manifold $I=J$ and $S=0$, it follows
that the number of angular momentum states associated with a
representation,  (2 J+1),  must equal the number of states in the
representation with $S=0$ (or equivalently with $Y= N_c /3$.

There is an ambiguity in how one implements this quantization. One
might choose to quantize the theory at large $N_c$ and then
systematically put in $1/N_c$ corrections. Alternatively, in
implementing the quantization condition  of eq.~(\ref{quantcond})
one can fix $N_c=3$ at the outset.  To the extent that $N_c=3$
can be considered large it ought not make any difference which of
these approaches is used, provided that one is studying states
which are not large $N_c$ artifacts.   Historically the choice of
taking $N_c=3$ at the outset has been standard\cite{SU3Quant}.
Making this choice, it is straightforward to see that the
lowest-lying states in this treatment are:
\begin{eqnarray}
J=1/2 \; \; \; (p,q) &=& (1,1) \; \; \;({\rm octet})  \nonumber \\
J=3/2 \; \; \; (p,q) &=& (3,0) \; \; \;({\rm decuplet})  \nonumber \\
J=1/2 \; \; \; (p,q) &=& (0,3) \; \; \;({\rm anti-decuplet}) \;
.\label{multi}
\end{eqnarray}
The decuplet and the anti-decuplet can then be seen to have  mass
splittings relative to the octet given by:
\begin{eqnarray}
M_{10} - M_{8} & = & \frac{3}{2 I_1} \; ,\label{10-8} \\
M_{\overline{10}} - M_{8} &  = & \frac{3}{2 I_2} \;  .
\label{10bar-8}
\end{eqnarray}

The preceding analysis is a variant of quite standard 1980's
vintage soliton physics. Note that this standard analysis of SU(3)
solitons is only justified in the large $N_c$ limit which plays an
essential role in two ways.  It justifies the use of the
classical static hedgehog configurations; effects of quantum
fluctuations around the hedgehogs are suppressed by $1/N_c$. It
also justifies the semi-classical treatment in collective
quantization; coupling between the collective motion and the
internal structure of the hedgehog is also suppressed by
$1/N_c$.  It should be clear from the previous comment,  however,
that the validity of the collective approach depends on
restricting its application to quantum collective modes. In order
to track the $N_c$ counting of various expressions we note that
the moments of inertia $I_{1,2}$  scale as $N_c$.

The regime of validity of collective motion is critical to the
analysis here, so it is useful to  specify what it is and where
it comes from.  The key point is that a collective description is
valid only for motion which is slow compared to the  vibrational
modes which are of order $N_c^0$.  The vibrational modes are
computed against a backdrop of a static soliton.  This is valid
providing the physical scale of the vibration is fast compared to
the scale over which the soliton rotates. If this is not true one
cannot separate the collective from the vibrational motion;  in
such a case the energy of the vibrational and collective motion
are not additive and, indeed, it is a misnomer to refer to it as
``collective'' motion.  Now the characteristic time scale of some
type of quantized collective motion is given by the typical
quantum mechanical result $\tau \sim (\Delta E)^{-1}$, where
$\Delta E$ is the splitting between two neighboring collective
levels.  Thus collective motion is valid only for motion for
which $\Delta E$ goes to zero in the large $N_c$ limit.

Conventional treatments of collectively quantized SU(3) solitons
identify the octet and decuplet states with the physical $N_c=3$
octets and decuplets familiar from baryon spectroscopy, while the
anti-decuplet has been dismissed as a large $N_c$ artifact in
much the same way that I=J=5/2 baryons are generally dismissed as
artifacts in SU(2) soliton models\cite{ANW}.  The principal
intellectual argument of ref.~\cite{DiaPetPoly} is that the
anti-decuplet should not be dismissed as a large $N_c$ artifact.
It argues that the anti-decuplet for SU(3) solitons can be
distinguished  from the J=I=5/2 baryons in SU(2) in an essential
way: the J=I=5/2 baryon width would be predicted to be so wide
with real world parameters that the state could not be
observed\cite{CohGri}. In contrast, the anti-decuplet state might
be expected to be narrow owing to suppressed phase space
associated with the increased mass of kaons relative to pions.
The fact that at the end of the calculation the predicted width
of the $\Theta^+$ is seen to be small is taken as a
self-consistent justification of this approach.

Before proceeding further, a brief remark about the calculation
in ref.~\cite{DiaPetPoly} is in order.  Much of the detailed
analysis concerns implementing SU(3) symmetry breaking effects in
the calculation and how to fit the resulting parameters from data.
For the present purposes, however, these are side issues. The
central question of principle is whether the predicted collective
anti-decuplet states are physical.

There is a very general argument why quantum number exotic
collective states in chiral soliton models are expected to be
spurious. A modern view of such models is that they encode the
predictions of large $N_c$ QCD relating the spin and flavor
dependence of various observables \cite{DJM}. The detailed
numbers emerging from the models---the values of the masses,
coupling constants and the like---are not reliable even at large
$N_c$ but the relations between them are. It is precisely because
the analysis of refs.~\cite{Pres,DiaPetPoly,WalKop,BorFabKob,Kim}
does not depend on dynamical details but merely on the structure
of the collective quantization,  that one might believe that it
correctly encodes the underlying QCD physics.  However, there is
an alternative method to deduce the spin-flavor properties of
large $N_c$ baryons in a model independent way via the use of
consistency conditions in describing meson-baryon
scattering\cite{DJM}. The results are well known: a contracted
SU(2$N_f$) symmetry emerges in the large $N_c$ limit.  Baryon
states fall into multiplets of SU(2$N_f$) and the low-lying states
in these multiplets are split from the ground state by energies
of order $1/N_c$---these excitations with the SU(2$N_f$)
multiplets are collective. Moreover, the multiplet of low-lying
baryons has been explicitly constructed---it coincides exactly
with the low spin states of a quark model with $N_c$ quarks
confined to a single s-wave orbital\cite{DJM}. Thus, it is well
known that there are no low-lying collective baryon states in
large $N_c$ QCD with quantum numbers which are exotic for the
large $N_c$ world.  In particular, there are no collective states
with strangeness +1 in large $N_c$ QCD. Any model which predicts
such a collective state appears to be inconsistent with large
$N_c$ QCD.

This general argument strongly suggests that any strangeness +1
state predicted via collective quantization of a chiral soliton
must somehow be spurious.  Yet, at first glance, the derivation of
eq.~(\ref{10bar-8}) appears to be based on standard chiral soliton
analysis.  The issue is what, if anything, is wrong with the
analysis?  The answer lies in the collective quantization.
Although the collective quantization of SU(3) solitons along the
lines of \cite{SU3Quant} is the standard for the field,
apparently, there has never been a careful study of the
conditions for which the approach is consistent with large $N_c$
QCD.  As will be shown below, the approach appears to give
excitations consistent with large $N_c$ QCD for the lowest-lying
$J=3/2$ states but {\it not} for the exotic strangeness +1 states.

As stressed previously, the standard semi-classical treatment for
collectively quantizing the solitons can only be justified in the
large $N_c$ approximation. The analysis outlined above appears to
respect the underlying large $N_c$ dynamics, at least formally.
After all, the mass splitting in eq.~(\ref{10bar-8}) goes as
$1/I_2 \sim 1/N_c$.  Thus, in the large $N_c$ limit the splitting
appears to become small which seems to imply that the motion is
collective. The semi-classical quantization approach thereby
looks to be justified self-consistently.

However, this is misleading: one can only see this collectivity
clearly in the large $N_c$ limit of the theory. Recall, however,
that eq.~(\ref{10bar-8}) was not derived in the large $N_c$
limit.  Its derivation depended on implementing  the quantization
condition in eq.~(\ref{quantcond}) with $N_c=3$ at the outset. It
was suggested above that making such a choice was innocuous, and
indeed it is, {\it provided the states being studied are not
artifacts}. However, since the entire question of relevance here
is whether the states are spurious, we cannot start by using
eq.~(\ref{10bar-8}) to see if the motion is truly collective.
Rather, one must study the full theory in its large $N_c$ limit
to see whether the motion turns out to be collective.

There are well-known peculiarities in studying SU(3) baryons in
the large $N_c$ limit.  First and foremost among these is the
fact that the SU(3) representations which emerge are not the ones
we are familiar with at $N_c=3$; indeed, as $N_c \rightarrow
\infty$ all of these SU(3) representations become infinite
dimensional\cite{DJM}.  However, this presents no insurmountable
problem phenomenologically, one simply associates those states in
the representation with isospin and strangeness quantum numbers
that survive down to the $N_c=3$ with their real world analogs.
The highly successful phenomenological study by Jenkins and Lebed
of baryon masses based on large $N_c$ scaling and SU(3) symmetry
and its breaking was based precisely on this
approach\cite{JenLeb}.

\begin{figure}
\includegraphics{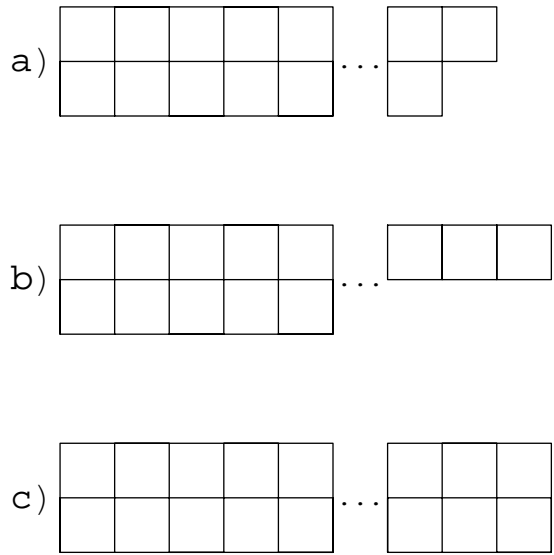}
\caption{ Young tableau for arbitrary but large $N_c$: a) the
``8'' representation with $(p,q)=\left( 1,\frac{N_c-1}{2} \right
)$; b)the ``10'' representation with $(p,q)=\left(
3,\frac{N_c-3}{2} \right )$: c) the ``$\overline{10}$''
representation with $(p,q)=\left( 0,\frac{N_c+3}{2} \right )$. The
young tableau in a) and b) have $N_c$ boxes;  the tableau in c)
has $N_c+3$ boxes   } \label{young}
\end{figure}

Consider the implementation of  eqs.~(\ref{quantcond}) and
(\ref{mass}) for $N_c$ arbitrary  and large.  To ensure that our
baryons remain fermions we  restrict our attention  to $N_c$ odd.
The lowest-lying representation compatible with
eq.~(\ref{quantcond}) is easily seen to be $ \left (p,q \right ) =
\left( 1, \frac{N_c-1}{2} \right) $ with $J=1/2$ and is
represented by the Young tableau a) in fig.~(\ref{young}).  The
states in this representation include those in the usual octet
(and are thus taken to be their large $N_c$ generalization); for
convenience this representation will be denoted ``8'' .  The
quotation marks serve to remind us that this is not really an
octet. The next representation is $ \left ( p,q \right ) = \left(
3,\frac{N_c-3}{2} \right ) $ with $J=3/2$; it is represented by
the Young tableau b) in fig.~(\ref{young}) and is denoted by
``10''. Using eq.~(\ref{mass}), it is straightforward to see that:
\begin{equation}
M_{ ``10"} - M_{``8"}  = \frac{3}{2 I_1}
\end{equation}
Note that this is identical to the analogous result for the
decuplet-octet splitting in eq.~(\ref{10-8}).  The significant
point, however, is that since $I_1$ scales as $N_c$, this
splitting {\it does} go to zero at large $N_c$ indicating that
the motion is, in fact, collective and thereby self-consistently
justifying the use of collective quantization.

Next consider a large $N_c$ representation analogous to the
$\overline{10}$.  The salient feature of the $\overline{10}$
representation is that it includes a state with strangeness +1.
Thus, its large $N_c$ analog should be taken to be the
lowest-lying representation that includes a state with
strangeness +1.  This representation is $ \left ( p,q \right ) =
\left( 0, \frac{N_c+3}{2} \right ) $ with $J=1/2$; it is
represented by the Young tableau c) in fig.~(\ref{young}) and is
denoted as ``$\overline{10}$''.  The excitation energy is given by
\begin{equation}
 M_{``\overline{10}"} - M_{``8"}  = \frac{3 + N_c}{4 I_2} \; .
 \label{10bar8quote}
\end{equation}
Of course,  eq.~(\ref{10bar8quote}) coincides with
eq.~(\ref{10bar-8}) for the special case of $N_c=3$.  However,
unlike eq.~(\ref{10bar-8}), eq.~(\ref{10bar8quote}) allows one to
study the $N_c$ scaling of the predicted splitting.  Note that
there is an explicit  $N_c$ in the numerator of the right-hand
side while the denominator is proportional to $I_2$ which scales
as $N_c$.  Thus, the scaling at large $N_c$ is given by
\begin{equation}
 M_{``\overline{10}"} - M_{``8"} \sim N_c^0 \; \; .
 \label{scaling}
 \end{equation}
In the large $N_c$ limit this splitting does not go to zero: the
excitation is {\it not} collective.  Note that the scaling in
eq.~(\ref{scaling}) is generic for states in large $N_c$ QCD
which are quantum number exotic in the sense that their quantum
numbers cannot be obtained from $N_c$ valance quarks.  It is
noteworthy  that the {\it only} states whose excitation energies
are of order $N_c^{-1}$ are those whose Young tableau contains
exactly $N_c$ boxes; these are precisely the one seen in the
general model independent analysis of ref. \cite{DJM}.

Recall that the energy of the exotic $\Theta^+$  was obtained
using the collective quantization {\it which is only valid for
collective modes}. However, as seen in eq.~(\ref{scaling}), it is
used to predict an excitation which is clearly not
collective---its excitation energy remains finite at large
$N_c$.  Thus, the prediction of the low-lying $\Theta^+$ state is
based on using collective quantization outside its domain of
validity.

Let us now revisit the argument in ref.~\cite{DiaPetPoly} based on
the predicted hadronic widths that the predicted anti-decuplet
state should not be regarded as spurious.  Note this argument
distinguished between the widths of the predicted anti-decuplet
and the J=5/2 states (which are generally regarded as large $N_c$
artifacts). From the perspective of this paper, it should be
clear that these two states are entirely different beasts.  The
$J=5/2$ states are collective modes whose properties one can
safely predict in a large $N_c$ world.  The sole issue  for the
predicted $J=5/2$ states is whether they survive in extrapolating
back from large $N_c$ to the real world at $N_c=3$.  In contrast,
the strangeness +1 exotic states are not collective even in the
large $N_c$ limit; treating them using collective quantization
will give rise to spuriously low energy modes.  In short, the
$J=5/2$ state is spurious because its prediction depends on taking
the large $N_c$ limit too seriously, while the collective
$\Theta^+$ state is spurious because its prediction depends on not
taking the large $N_c$ limit seriously enough.   Thus, although
the reasons for which one regards the J=5/2 state as spurious do
not apply to the anti-decuplet, the anti-decuplet is spurious for
entirely different reasons.

In summary, the predicted $\Theta^+$ baryon in
ref.~\cite{Pres,DiaPetPoly,WalKop,BorFabKob,Kim} was obtained
using collective quantization in a regime where collective
quantization does not apply.  It was shown that quantum number
exotic states in large $N_c$ QCD have excitation energies which
are of order $N_c^0$ and thus are not collective. Accordingly,
the prediction of the $\Theta^+$ as a collective excitation
should be regarded as being invalid; the fact that the predicted
mass was so near to the observed mass must be regarded as
fortuitous.

Of course, none of the arguments presented here indicate that
chiral soliton models are intrinsically incapable of describing
exotic states or indeed of doing a reasonable phenomenological
job in describing the $\Theta^+$ baryon. However, if exotic
states do exist in this class of models, they must be obtained by
methods which are suitable to describe excitations of order
$N_c^0$ rather than $N_c^{-1}$. Such methods do exist.  For
example one can use linear response theory to describe mesons
scattering from baryons\cite{KarMat}. In principle, an exotic
$\Theta^+$ state could emerge in such a picture as a resonant
state of a kaon and an ordinary baryon. However, there is no
general argument that an exotic resonance would be generated for
all such models and the excitation energy of such a state, if it
exists, is completely model dependent.  This does not imply that
such an analysis is useless.  One important aspect of large $N_c$
QCD is that it {\it correlates} predictions.  In particular, the
existence of one light strangeness +1 resonant state implies the
existence of other strangeness +1 resonant states which differ in
energy from it by of order $1/N_c$.  While the arguments
presented in this paper show why the order $N_c^0$ splitting
between the ground state and the exotic are unreliable, the order
$1/N_c$ splittings between exotic states are reliable. These
predicted new states are explored in ref.~\cite{CohLebTheta}.

The author acknowledges Rich Lebed for several insightful remarks
about this work.  The support of the U.S. Department of Energy
for this research under grant DE-FG02-93ER-40762 is gratefully
acknowledged.


\begin{thebibliography}{99}
\bibitem{Nus} It has been argued, however, that the width's reported in
the experiments may be broadened by experimental issues
associated with resolution.  Comparison with previous data seems
to suggest that the actual width may be much narrower.  This
argument is detailed in S. Nussinov,  hep-ph/0307357.
\end{thebibliography}
\end{document}